\documentclass[11pt,a4paper]{article}
\usepackage{graphicx}
\begin{document}
\thispagestyle{empty}
\renewcommand{\baselinestretch}{1.2}
\small\normalsize
\frenchspacing
\noindent
{\Large \textbf{Picturing wavepacket reduction}}
\\
\\
{\bf{Arthur Jabs}}
\\

\renewcommand{\baselinestretch}{1.05}
\small\normalsize
\noindent
Alumnus, Technical University Berlin.\\ 
Vo\ss str. 9, 10117 Berlin, Germany

\noindent
arthur.jabs@alumni.tu-berlin.de
\\ \\
\noindent
(3 Oct 2017)
\newcommand{\rmi}{\mathrm{i}}
\bigskip

\noindent
{\bf{Abstract.}} A coherent picture of the wavepacket-reduction process is proposed which is formulated in the framework of a deterministic and realist interpretation where the concepts of knowledge or information and of point particles do not appear. It is shown how the picture accounts for  the experiments on interaction-free and delayed-choice measurements and those on interference with partial absorption.

\vspace{-2pt}
\begin{list}
{\textbf{Keywords:}}
{\setlength{\labelwidth}{2.0cm}
 \setlength{\leftmargin}{2.2cm}
 \setlength{\labelsep}{0.2cm} }
\item
reduction, collapse, delayed choice, interaction-free measurements
\end{list}
\vspace{5pt}
\\
\rule{\textwidth}{.3pt}
\newcommand{\leer}{\hspace*{\fill}}
\newcommand{\lll}{\hspace*{90pt}}
\newcommand{\hsp}{\hspace*{60pt}}
\begin{center}
\vspace{-20pt}
\noindent
 \hsp 1~~ Realism \leer 1\hsp \lll \\
\hsp 2~~ Reduction. Criterion \leer 2\hsp \lll \\
 \hsp 3~~ Reduction. Complement \leer 3\hsp \lll \\
 \hsp 4~~ Interaction-free measurements. Theory \leer 5\hsp \lll \\
\hsp 5~~ Interaction-free measurements. Experiment \leer 6\hsp \lll \\
 \hsp 6~~ Delayed-choice experiments \leer 9\hsp \lll \\
\hsp 7~~ Partial absorption \leer 11\hsp \lll \\
 \hsp Notes and references	\leer 12-15\hsp \lll \\
\end{center}
\vspace{-10pt}
\rule{\textwidth}{.3pt}
\\

\newcommand{\rmf}{\mathrm{f}}
\newcommand{\rmd}{\mathrm{d}} 
\newcommand{\sbl}{\hspace{1pt}}
\newcommand{\bfitp}{\emph{\boldmath $p$}}
\newcommand{\bfitr}{\emph{\boldmath $r$}}
\newcommand{\bfitsr}{\emph{\footnotesize{\boldmath $r$}}}
\newcommand{\bfitQ}{\emph{\boldmath $Q$}}
\newcommand{\bfitk}{\emph{\boldmath $k$}}
\newcommand{\PSI}[1]{\Psi_{\textrm{\footnotesize{#1}}}}
\newcommand{\PHI}[1]{\Phi_{\textrm{\footnotesize{#1}}}}
\newcommand{\spsi}[1]{\psi_{\textrm{\footnotesize{#1}}}}
\newcommand{\pcop}{P_{\textrm{\footnotesize{Cop}}}}
\newcommand{\pred}{P_{\textrm{\footnotesize{red}}}}
\newcommand{\psis}{\psi_{\rm s}(\bfitr,t)}
\vspace{8pt}

\bigskip

\noindent
{\textbf{1~~Realism}}
\smallskip

\noindent
The picture of the reduction process (collapse) presented in this note is based on the concepts introduced in an interpretation of the quantum-mechanical formalism in terms of epistemological realism [1] and in a conjecture of how determinism can be introduced into quantum mechanics [2].  In complementing the conjecture made in [2] the picture  gives a coherent account of such physical situations, both with massive particles and with photons in the spirit of Einstein [3], which are particularly incomprehensible in terms of the traditional concepts.

In this section a brief account is given of those features of [1] which are relevant for the proposed picture. The gross features of reduction and the deterministic criterion for it, according to [2], will be described in the next section. In Sec.~3 the picture is complemented by an additional feature covering a situation which was not treated in [2]. In Sec.~4  Elitzur and Vaidman's thought experiment on interaction-free measurements [4] is interpreted, and in Sec.~5 experimental realizations of that experiment are discussed. In Sec.~6 a consistent interpretation of Wheeler's delayed-choice thought experiment and its experimental realizations is given, and in Sec.~7 the case of partial absorption in the experiment by Summhammer, Rauch, and Tuppinger [5] is discussed.
 
In the realist interpretation [1] the wave functions are not probability amplitudes for outcomes of measurements but real physical objects, in the sense that pulses of electromagnetic radiation are real objects. A ``one-particle'' (one-quantum) wave function does not contain and guide a pointlike particle but itself represents the quantum object. There is no ``wave-particle duality'' but there are only waves, albeit with special properties. The word ``particle'', if we continue using this word for the quantum object, acquires a new meaning, namely that of a wavepacket of finite extension, where even spatially separated (``entangled'') parts of the packet can represent one and the same quantum object. The poinlike appearance of the extended wavepacket in a measurement comes from its contraction to a small region in the reduction process, as will be explained below.

The question of whether the Heisenberg relations can or cannot be ascribed to the action of a measurement is meaningless here because they are inherent properties of wavepackets namely Fourier reciprocity relations, already well known for pulses of classical electromagnetic radiation [1, Sec. 2.5] (cf. also [6]).

To account for typical quantum effects such a wavepacket is endowed with the special property of nonlocality.  With this property the packet can give rise to correlated spacelike events which are not completely determined by common causes. If the wavepacket contracts at a particular place, the nonlocality means that it contracts with superluminal speed [7]. 

The second type of quantum object with this property is a system of entangled wavepackets [1, Sec. 3.1, entanglement distribution or swapping]. If one wavepacket of a pair of entangled packets is reduced, the other one is reduced too [8], and the two reduction events occur at spacelike distances [9] (cf. Sec. 6 below).

\vspace{10pt}
\noindent
\textbf{2~~Reduction. Criterion} 
\smallskip

\noindent
Consider a one-quantum (one-particle, one-photon) wavepacket which approaches a screen oriented perpendicularly with respect to the direction of propagation. The screen is a macroscopic body with small sensitive clusters in it. A cluster is a group of strongly bound atoms or molecules, or only a single atom. To be of use in a measurement the clusters must be small compared with all other relevant dimensions. The incident  `measured' wavepacket, when it gets in contact with one of these clusters, may then become reduced, that is, contracted to the size of the cluster [2]. 

Reduction/contraction is independent of measurement, but measurement needs it. As conjectured in [2] the criterion for the contraction of the incoming wavepacket to occur at some time and at a cluster located about $\bfitr_0$ consists of two conditions:
\[
\hspace{155pt}|\alpha_1-\alpha_2|\leq\frac{1}{2}\alpha_{\textrm{s}} \hspace{150pt}(1)
\]
\vspace{-15pt}
and
\[
\hspace{122pt}  \bigg[\int_{\rm R^3} |\spsi{i}(\bfitr,t)| \, |\spsi{cl}(\bfitr,t)| \, {\textrm d}^3r\bigg]^2 \ge \alpha/{2\pi}. \hspace{80pt}(2)
\]
$\alpha_1$ is the absolute (overall, global) phase constant of the incoming wavepacket $\spsi{i}(\bfitr,t)$, and $\alpha_2$ that of the wavepacket    $\spsi{cl}(\bfitr,t)$  which represents the cluster. $\alpha_{\textrm{s}}=e^2/\hbar c \approx  1/137$ is Sommerfeld's fine-structure constant.  

Given that the phase constants are such that the first (the phase-matching)  condition (1) is satisfied for a particular cluster, the contraction occurs at the first moment $t$ when the second (the overlap) condition (2) is also satisfied (if ever). 

In a measurement the clusters met by the incident wavepacket must be of a special kind, namely such that an avalanche of processes originates from them and leads to an observable macroscopic effect, a ``spot''. In [2] it is shown how the criterion (1), (2) leads to the standard Born rules when the absolute phase constants are regarded as pseudorandom numbers uniformly distributed in $[0,2\pi]$. 

The exact form of the criterion is not important for the present considerations. What is essential is that the one-quantum wavepacket can undergo only one reduction at a time, that reduction  means contraction to the size of a cluster, and that some  nonlocal ``hidden'' variables (in our approach the absolute phase constants) determine whether or not a contraction occurs at $t$ and $\bfitr_0$.

A one-photon wavepacket is absorbed and vanishes upon reduction, but actually does so at one single narrow place, so that we speak of reduction as meaning contraction both for massive and massless wavepackets.

\vspace{10pt}
\noindent
{\textbf{3~~Reduction. Complement}}
\smallskip

\noindent
Ref.~[2] was focussed on situations where at a particular cluster both conditions of the criterion say Yes, and reduction leads to a spot at that cluster. In this note we consider physical situations where the first condition says Yes, but the second says No.

Consider the situation sketched in Fig.~1a: a beam of wavepackets (photons, neutrons, atoms) meets a 50:50 beam splitter BS where it is split in two equal branches, $a$ and $b$. The branches end up in two equal detectors D1 and D2, respectively, which both have the same distance from the beam splitter BS. In this situation the number of clicks in D1 in the long run equals the number in D2.

Let the intensity of the beam be so low that never more than one  wavepacket at a time is in the region between BS and D1 or D2. Then each single wavepacket is split in two equal parts which have definite phase relations and together represent one and the same quantum wavepacket.

The clicks of the detectors are connected with reduction, that is with contraction at one of the clusters in D1 or in D2.  Of course, owing to the fact that the wavepacket contracts as a whole and to a narrow place, never more than one detector can click at a time (e.g. [10]). 

We meet however a difficulty with the picture developed so far when we consider the modification of the situation in Fig.~1a which is sketched in Fig.~1b: detector D1 is replaced by detector AD1 in which the number of sensitive clusters is largely enhanced so that AD1 would be a complete absorber, that is, no wavepacket which enters it could escape it. Detector D2 remains unchanged, but is put many kilometers away from the beam splitter BS.

\vspace{15pt}
\begin{figure}[h]
\begin{center}
\includegraphics[width=0.75\textwidth]{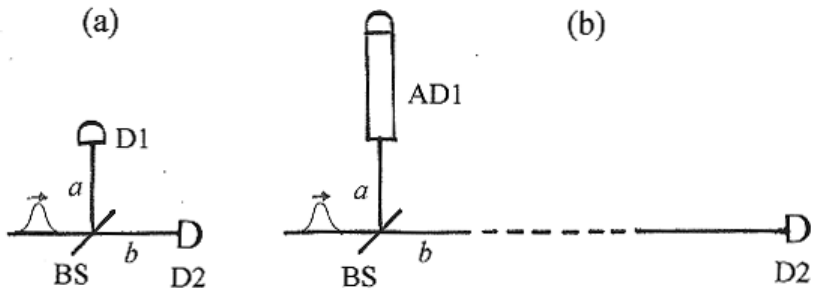}
\caption{(a) basic and (b) modified physical situation}
\end{center}
\end{figure}

Now, our considerations are based on the supposition that even in this situation in half of the cases detector AD1 clicks and part $b$ disappears, and in the other half detector D2 clicks and part $a$ disappears, this being \emph{independent of the distance between beam splitter} BS \emph{and detector} D2. This is what the formulas of standard quantum mechanics predict and what can be explained in a point-particle picture. The above supposition is confirmed to a certain degree in the experiments on ``interaction-free'' measurements considered in Secs.~4 and 5, where the distance which corresponds to BS--D2 is larger that the distance BS--AD1 though not very much larger.

In a pure wavepacket picture the question arises: why is the wavepacket not in all observed cases, rather in only half of them, contracted and absorbed in the nearby absorbing detector AD1? The contraction occurs when the incoming wavepacket on its way with part $a$ meets a cluster in AD1 that satisfies the phase-matching condition (1) and also the condition (2). Now, the absorbing detector AD1 contains so many clusters that there is with certainty one cluster which satisfies both conditions of the criterion and makes part $b$ of the wavepacket vanish long before it could reach the distant detector D2. Therefore, to avoid this, we  propose to endow the reduction process with an additional feature:

If at the first cluster in AD1 for which the phase-matching condition (1) is satisfied, condition (2) allows a contraction, then part $a$ is contracted and part $b$ vanishes.

 Conversely, and this is the case which goes beyond [2], if (1) is satisfied but (2) forbids a contraction, \emph{part $a$ vanishes and part $b$ survives} with unchanged form but normalization raised up to 1, that is, the remaining parts of the wavepacket are multiplied with a common factor. For this to occur we conjecture generally that between part $a$ and part $b$ at the moment of the first contact with a sensitive cluster (i.e. where (1) says Yes) 
\begin{quote}
\emph{the parts of the wavepacket must lie in disjoint regions of phase space.}
\end{quote}
This is a property which is invariant under Lorentz transformations. We want to call it phase-space separation.

In a situation where there is no part which is phase-space separated from part $a$, say, then part $a$ could indeed undergo repeated contractions in AD1 (except one-photon packets), even when in the first phase-matching clusters condition (2) has said No. Such a situation could, for example, be realized when the beam splitter in Fig.~1b is replaced by a mirror, or when elementary particles leave tracks in nuclear emulsion or cloud chambers. 

Conversely, the absorber may be so short that the wavepacket in traversing it would only rarely meet a phase-matching cluster. That would then mean a partial absorber, considered in Sec.~7, when we disregard absorption due to the Schr\"odinger evolution.

\vspace{10pt}
\noindent
{\textbf{4~~Interaction-free measurements. Theory}}
\smallskip

\noindent
We are now going to consider coincidences between detectors at different distances from the beam splitter, or of detector clicks in situations where one part of the wavepacket is blocked before the other part can reach a detector. We shall see that the picture described  in the preceding sections is capable of giving a consistent account of the results.

We begin by considering the ``interaction-free'' measurement discussed as a  thought experiment by Elitzur and Vaidman [4]. They consider the situation shown in Fig.~2. A one-quantum wavepacket reaches the first beam splitter BS1 where it is split in two equal parts, $a$ and $b$. These are then reflected by the mirrors in such a way that they are reunited at another, similar beam splitter BS2. Two detectors, D1 and D2, can detect the parts with 100 \% efficiency. The positions of the beam splitters and the mirrors are arranged so that, due to destructive interference, detector D2 never, but D1 always clicks. Now an impenetrable macroscopic object O may or may not be put in branch $a$. The distance BS1--D1 (or BS1--D2) is larger than the distance BS1--O, and this corresponds to the situation described in Sec.~3 and sketched in Fig. 1(b). From the then observed detector clicks one may find out whether the object is actually there.
\vspace{7pt}
\begin{figure}[h]
\begin{center}
\includegraphics[width=0.7\textwidth]{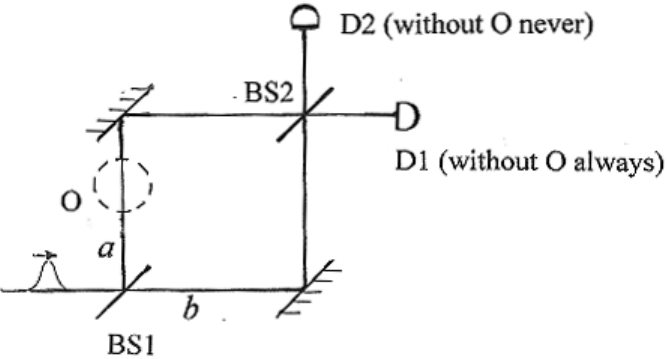}
\caption{The situation considered by Elitzur and Vaidman [4]}
\end{center}
\end{figure}

By looking at Fig.~2 one sees that there are three possible outcomes of this measurement: (i) if detector D1 clicks (probability 1/4) the object may or may not be there and one can say nothing, (ii) if no detector clicks (probability 1/2) the object is there and has absorbed the particle, and (iii) if detector D2 clicks (probability 1/4) the object is there, but has not absorbed the particle.

Case (iii) is the case which Elitzur and Vaidman call an interaction-free measurement: the measurement confirms the presence of the object though the particle has not interacted with it. In Elitzur and Vaidman's situation our question is: why is the wavepacket not always contracted and absorbed in the object O so that nothing reaches either detector D1 or D2? As discussed in Sec.~3 our answer is that in this case the overlap condition (2) forbids contraction at the very first phase-matching cluster in O so that part $a$ vanishes and part $b$ survives and reaches detector D1.

In the considered case of equal parts $a$ and $b$ the ratio $\eta$ of the probabilities of detecting the object without absorption to that with absorption is
\vspace{-1pt}
\[
\eta=\frac{P(\rm detector\; {\rm D2}\; clicks)}{P(\rm no\; detector\; clicks)}=\frac{1/4}{1/2}=\frac{1}{2}.
\]
As demonstrated in [4] this ratio can be increased to nearly 1 if beam splitter BS1 is almost completely transparent and BS2 almost completely reflecting. Part $b$ is then much larger than part $a$, and D1 almost always clicks. A disadvantage of this is that it is the very case in which no conclusion on absence or presence of the object can be drawn.
In our picture the probability that part $a$ then meets a phase-matching cluster in the object is unchanged because all parts of the wavepacket have the same absolute phase constant. The overlap condition (2) is however almost never satisfied for part $a$, so that almost always part $b$ takes over and D1 clicks.

\vspace{10pt}
\noindent
{\textbf{5~~Interaction-free measurements. Experiment}}
\nopagebreak
\smallskip

\noindent
Elitzur and Vaidman's thought experiment has been experimentally realized in [11] - [18], and the results confirm the predictions. Our explanation of the results only works if only one wavepacket at a time is in the apparatus and the two parts of the wavepacket, at the moment when part $a$ reaches the object O, are phase-space separated. We therefore have to check whether these conditions are satisfied in the experiments. Indeed we found that there was only one wavepacket at a time in the respective apparatuses. Separation in momentum space between the parts of the wavepacket is easy to see when the two parts move in different directions. Checking the separation in ordinary space was not always an easy task because the data which describe the experimental setup, in particular the spatial distances between the various components of the apparatuses and the length of the wavepackets used, are not always reported and one has to resort to other papers which describe analogous situations. In particular, the question of wavepacket length deserves special attention:

In [11] neutrons were sent through the perfect silicon crystal neutron interferometer. Consulting [19] we concluded that the coherence length $\sigma_{\textrm{\footnotesize{cy}}}$ of the incident neutron beam was $10^{-8}$ m, which is indeed very short compared with the distances of about 5 cm between the components of the interferometer. However, it is not the coherence length $\sigma_{\textrm{\footnotesize{cy}}}$ that is responsible for detector clicks, but this is the total length $\sigma_{\textrm{\footnotesize{y}}}$ of the wavepacket. While $\sigma_{\textrm{\footnotesize{cy}}}$ is constant in time, $\sigma_{\textrm{\footnotesize{y}}}$ spreads out and may become much larger than $\sigma_{\textrm{\footnotesize{cy}}}$ [20], [21]. A rough though acceptable estimate of longitudinal spreading is the following: we may write $\sigma_{\textrm{\footnotesize{y}}} =  \sigma_{\textrm{\footnotesize{cy}}} + \Delta_{\textrm{\footnotesize{sy}}}$, where  $\Delta_{\textrm{\footnotesize{sy}}}$ is the increase due to spreading. For a nonrelativistic wavepacket [1, Appendix A] it is $\Delta_{\textrm{\footnotesize{sy}}}\approx l\,\Delta \lambda/\lambda$, and with $l=5$ cm, $\Delta \lambda/\lambda=0.01$, it is $\Delta_{\textrm{\footnotesize{sy}}}=0.5\times10^{-3}$ m. This is large compared with $\sigma_{\textrm{\footnotesize{cy}}}=10^{-8}$ m, so that  
$\sigma_{\textrm{\footnotesize{y}}}\approx \Delta_{\textrm{\footnotesize{sy}}}=0.5$ mm, but  even after spreading the length $\sigma_{\textrm{\footnotesize{y}}}$ of the wavepacket, and thus also of its parts, is small compared with the distance $l=5$ cm, and this means a complete separation in ordinary space.

Refs. [12] - [18] deal with photons. For photons there is no longitudinal spreading [1, Appendix~A]. In many experiments the incident photon packet was one of a pair  from parametric down-conversion. These packets have a length of about $30\,\mu$m [22], [23],  and this is short compared with the dimensions of the apparatus used.

Indeed, in all experiments we found that phase-space separation was realized.
 
The paper [17] suggests a side remark illustrating clearly the general feature of treating continuous intensities as frequencies of discrete events when going from the classical to the quantum domain. The principle of this experiment is sketched in Fig.~3. 
\begin{figure}[h]
\begin{center}
\includegraphics[width=0.6\textwidth]{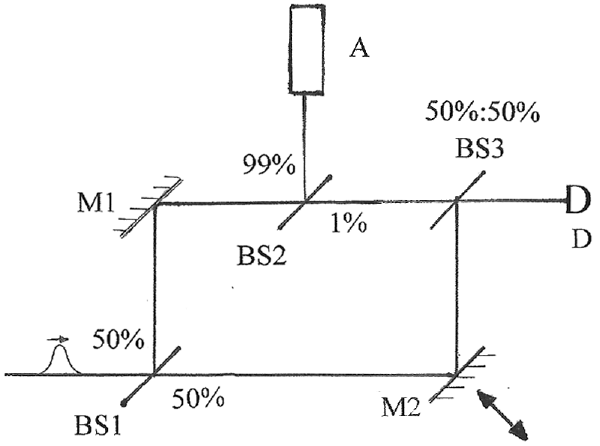}
\caption{Principle of the experiment in [17]: demonstration of classical versus quantum visibility}
\end{center}
\end{figure}
A is a complete absorber. Variations in the intensity due to interferences, produced by moving the mirrors, or by other means, are observed with detector D. The beam consists of wavepackets and each packet is split in two very different parts at the asymmetric 1:99 beam splitter BS2. These parts are recombined and interfere at the 50:50 beam splitter BS3.

In classical electrodynamics the two interfering parts have very different intensities at D, namely  $(0.5\times0.5)$ and $(0.5\times0.5\times0.01)$, so that the visibility of the interference fringes is $V:=(I_{\textrm{\footnotesize{max}}}-I_{\textrm{\footnotesize{min}}})/(I_{\textrm{\footnotesize{max}}}+I_{\textrm{\footnotesize{min}}})=0.01$.
 
With quantum wavepackets it is different: In 99 \% of the cases the whole packet is absorbed (reduced), so that no packet or part of a packet arrives at D. In 1 \% of the cases the packet passes beam splitter BS2 as if the splitter would not be there and the photon can arrive at D. The visibility of the \emph{recorded} photons thus is $V=1$ because the interfering parts have the same amplitude, and  $I_{\textrm{\footnotesize{max}}}=1$ and $I_{\textrm{\footnotesize{min}}}=0$ at D. In the experiment [17] with photon packets a value of $V=0.86$ was obtained. The deviation from the value 1 is ascribed to imperfections of the apparatus.

Finally, are the ``interaction-free'' measurements really interaction free? The term was first used by Dicke [24].

In our picture the votes against that denomination are that there must be some contact of the incoming wavepacket with the wavepacket that represents the sensitive cluster in the measuring apparatus in order that phase matching according to condition (1) can be established. Even if the second condition, (2), then decides that no contraction will occur, the incoming wavepacket, provided it consists of several phase-space separated parts, does not go on by Schr\"odinger dynamics, but is modified in that the part of it which got into contact with the cluster vanishes in favour of its other, separated parts. -- Also, in the case of a photon wavepacket there is no absorption of the photon in case (iii) of Sec.~4, but there are other processes which do not absorb the photon and are yet called interactions: reflexion (a moving mirror even changes the energy), refraction, quartz wave plates, $\lambda$/4-plates etc. [25], [26].

Votes in favour of the denomination are that the contact which establishes phase matching and leads to contraction is a process which is not governed by a Schr\"odinger equation with a term called interaction term in the Hamilton operator, but is a process of a different  kind.

\vspace{10pt}
\noindent
{\textbf{6~~Delayed-choice experiments}}
\smallskip

\noindent
In Wheeler's delayed-choice thought experiment [27], [28] the situation already depicted in Fig.~2 is considered, but without the object O and with a beam splitter BS2 that could be introduced or removed. In a na\"ive point-particle picture the incoming particle should make its choice of taking either path $a$ or path $b$ at the 50:50 beam splitter BS1. This seems to be confirmed by the observation that when the beam splitter BS2 is removed either detector D1 or detector D2 clicks. However, when BS2 is introduced it can be arranged, by means of constructive and destructive interference between waves from the two paths, that always detector D1, say, clicks, and never detector D2. The experimenter's decision to introduce or remove BS2 can be taken after the particle has passed the input beam splitter BS1, and shortly before it reaches the location of BS2. Thus, the particle must have delayed its choice of whether going along path $a$ or along path $b$ or along both paths simultaneously, that is between which-path or both-path behaviour, to a moment well after it has passed the input beam splitter BS1. The choice is also regarded as one between particle- and wave-like behaviour.

The paradox is, in fact, an artifact of the point-particle interpretation of quantum mechanics. It does not appear in the wavepacket interpretation advanced in the present note: what enters the interferometer is a wavepacket, and this is split in the two equal parts, $a$ and $b$ of Fig.~2. If the beam splitter BS2 is removed, part $a$ arrives at detector D1 and part $b$ at D2. Which detector will click depends on which is the first in which the respective part meets a phase-matching cluster and satisfies condition (2). The wavepacket then contracts to the size of that cluster, and due to the nonlocality of the quantum wavepackets, at that same moment the other part of the wavepacket, which was on the way to the other detector, vanishes. According to what we have said in Sec.~3 this is independent of whether the wavepacket and with it its parts at this moment are longer or shorter than the distance between BS2 and D1 or D2. If the wavepacket is longer and the ways from BS1 to D1 and D2 have the same length the situation is like the wavepacket approaching a screen. If the wavepacket is shorter and in one detector condition (1) but not condition (2) is satisfied with one part of the wavepacket, then the other part takes over. Of course, never more than one wavepacket at a time should be in the apparatus.

\vspace{5pt}
There are many experimental realizations of Wheeler's thought experiment. Ref.~[29] is done with neutrons, [30] with atoms, and [31] - [34] with photons.

The necessary conditions for the application of our explanation, as described above and in Sec.~3, have been checked and found to be satisfied in these experiments. In [31] the two events: the wavepacket's  passage at BS1 and the decision to insert or remove BS2 are even separated by a spacelike distance in spacetime, accomplished by a quantum random number generator, so that even a light signal from BS1 could not have influenced the decision at BS2.

It is also interesting to consider those delayed-choice experiments where the choice concerning the quantum wavepacket under consideration, call it $\psi_{\rm B}$, is effected by a second packet, $\psi_{\rm A}$, which is entangled with   $\psi_{\rm B}$  . Packet   $\psi_{\rm A}$   enters a measuring apparatus, which, by means of reduction, turns $\psi_{\rm A}$ into one of two possible components. Due to the entanglement  $\psi_{\rm B}$ is turned into the corresponding  component, which is then directed into an apparatus which allows one component to produce interferences (``wave-like'' behaviour), but does not allow this for the other component (``particle-like'' behaviour). Experiments of this kind are reported in [35] - [38].

As a concrete example let us consider the basic features of the experiment in [35], which was done with a pair of entangled photons. The two superposed components considered of the photons are here those with horizontal and vertical linear polarization. Photon $\psi_{\rm B}$ enters Bob's Mach-Zehnder interferometer, similar to the one depicted in Fig. 2, but without the object O. BS1 is the usual 50:50 beam splitter and splits each component in two parts going path $a$ and path $b$, respectively. BS2, however, now is polarization dependent, that is, it is a 50:50 splitter for the vertical component, but no splitter at all for the horizontal component. Thus  the two parts of the vertical component can be superposed, and by phase shifting one part, interference effects can be observed because the counting rates in detector D1 or D2 will depend on the phase shift (``wave-like'' behaviour). All detectors consist of a polarizing beam splitter followed by two counters, C$_{\rm H}$ for the horizontal, and C$_{\rm V}$ for the vertical component.
The two parts of the horizontal component, however, pass BS2 without influencing each other, so that either detector D1 is made to click by part $a$ or D2 by part $b$, and no interference effects can be observed (``particle-like'' behaviour).

The other photon packet, $\psi_{\rm A}$, is directed to Alice's Mach-Zehnder interferometer, which is equal to Bob's. Thus $\psi_{\rm A}$ is also directed to a detector whose two counters, C$_{\rm V}$ and C$_{\rm H}$, click for the vertical or the horizontal component, respectively. Thus if C$_{\rm V}$ clicks, $\psi_{\rm A}$ is reduced to its vertical component and vanishes, and due to the entanglement Bob's photon packet  $\psi_{\rm B}$ is also reduced to the vertical component, which persists. This component can then produce interference effects and show ``wave-like'' behaviour. If, on the contrary, in Alice's detector the counter C$_{\rm H}$ clicks, her photon is reduced to the horizontal component, Bob's photon too, and Bob cannot obtain interferences. In this way Alice's photon governs the choice between particle- or wave-like behaviour of Bob's photon.

In our picture the reduction events of $\psi_{\rm A}$ and $\psi_{\rm B}$ occur simultaneously in some refernce system, that is, at spacelike separation in spacetime. Thus, if in the laboratory frame Alice's detectors click first and Bob's later, in another frame it may be the reverse. In a frame where Bob's detectors click first, Bob's photon packet $\psi_{\rm B}$ and thus also Alice's packet $\psi_{\rm A}$ are reduced to their respective components. But remember that neither Alice nor Bob by themselves can determine which one of their counters (horizontal or vertical) will click, and which click is the cause and which the effect. This forbids superluminal signaling, and there is no paradox in the picture proposed here.

\vspace{10pt}
\noindent
{\textbf{7~~Partial absorption}}
\smallskip

\noindent
We will condider here a special aspect of the neutron-interference experiment by Summhammer, Rauch, and Tuppinger [5]. In this experiment an incoming short neutron wavepacket $\psi$ is split in two parts, $\psi_{\rm L}$ and $\psi_{\rm R}$, with equal intensities, and a partially absorbing foil with transmission probability $a$ is inserted in the path of $\psi_{\rm L}$. The effect of this absorber was not that in the fraction $a$ of the incoming wavepackets the packet passes the $\psi_{\rm A}$  absorber unaffected and in the fraction $(1-a)$ the packet with all its parts is completely blocked (as happens with a chopper of the same transmission probability). 

This could be seen when the escaping part of $\psi_{\rm L}$ was brought to interference with $\psi_{\rm R}$: the part $\psi_{\rm L}$ which entered the absorber escaped it as $\psi_{\rm L}^{\prime}=\sqrt{a}\psi_{\rm L}$ and interferes as such with $\psi_{\rm R}$. The intensity of the superposition of $\sqrt{a}\psi_{\rm L}$ and the phase-shifted $\psi_{\rm R}^{\prime}=\exp(\rmi\phi)\psi_{\rm R}$ is 
\begin{displaymath}
\hspace{45pt}
I(\phi)=|\sqrt{a}\psi_{\rm L}+\exp(\rmi\phi)\psi_{\rm R}|^2=|\psi_{\rm L}|^2(1+a+2\sqrt{a}\cos(\phi)) 
\hspace{60pt}(3)
\end{displaymath}
because at the detector we may put $\psi_{\rm L}=\psi_{\rm R}$. From this the normalized amplitude of the interference pattern  \hspace{3pt} $A_{\rm n}$=$(I_{\rm max}-I_{\rm min})/(I^0_{\rm max}-I^0_{\rm min})=\sqrt{a}$  \hspace{3pt} is obtained, where $I^0_{\rm max}$ and $I^0_{\rm min}$ are the intensities when the transmission probability is $a$=1. This $\sqrt{a}$ dependence of $A_{\rm n}$ (as opposed to an $a$ dependence with a chopper) was observed in [5] as well in analogous experiments with photons [39], [40].

We want to show that even when the contraction, as described in Secs.~2 and 3, is taken into account, the dependence $A_{\rm n}$=$\sqrt{a}$ obtains. For this we will have to assume that $\psi_{\rm L}$, when entering the absorber is in turn split in two different parts, $\psi_{{\rm L}1}$ and $\psi_{{\rm L}2}$, with $|\psi_{{\rm L}1}|^2=a|\psi_{\rm L}|^2$ and $|\psi_{{\rm L}2}|^2=(1-a)|\psi_{\rm L}|^2$. This is in any case necessary if we want to stick to a constant overall normalization of the incoming wavepacket $\psi$ in any stage (split or non-split) within the interferometer. The part $\psi_{\rm L1}$, say, continues on the way to the final detector and can be written as $\psi_{\rm L1}=\sqrt{a}\psi _{\rm L}$, but $\psi_{\rm L2}$ is deflected and can never reach the detector. These two parts are thus phase-space separated. A splitting of this kind is also taken into account in de Muynck et al.'s theoretical treatment [41] of Summhammer et al.'s experiment [5].

Now, in our picture, in the fraction $q_1$ of the incoming wavepackets let either $\psi_{{\rm L}1}$ or $\psi_{{\rm L}2}$ meet a phase-matching cluster in the absorber. If in such a case condition (2) says Yes, the whole wavepacket, with all its parts, vanishes, and nothing reaches the detector.

If, on the other hand, condition (2) says No, then either $\psi_{{\rm L}1}$ at the cost of $\psi_{{\rm L}2}$, or $\psi_{{\rm L}2}$ at the cost of $\psi_{{\rm L}1}$, takes over and the remaining parts of the wavepacket, namely either $\psi_{\rm R}$ and $\psi_{{\rm L}1}$, or $\psi_{\rm R}$ and $\psi_{{\rm L}2}$, are multiplied by a common factor so as to maintain the overall normalization.

Now, if in this subcase $\psi_{{\rm L}2}$ takes over, then $\psi_{{\rm L}1}$ vanishes, and there is nothing to interfere with $\psi_{\rm R}$. But if $\psi_{{\rm L}1}$ takes over and $\psi_{{\rm L}2}$ vanishes, $\psi_{{\rm L}1}$ and $\psi_{\rm R}$ are multiplied by a certain common factor $f$. {\rm L}et this happen in the fraction $q_2$ of the neutrons which meet a phase-matching cluster. The modified $\psi_{\rm L}$ and $\psi_{\rm R}$ can then be brought to interference and the intensity of their superposition is $I^{\prime}(\phi)=q_1q_2|f\sqrt{a}\psi_{\rm L}+f\exp(\rmi\phi)\psi_{\rm R}|^2=q_1q_2f^2I(\phi)$ and by using Eq.(3) we are again led to $A_{\rm n}=\sqrt{a}$, which is what we wanted to show.

\vspace{10pt}
\noindent
{\textbf{Notes and References}}
\begin{enumerate}
\renewcommand{\labelenumi}{[\arabic{enumi}]}

\item Jabs, A.: Quantum Mechanics in Terms of Realism, arXiv:quant-ph/9606017 (2017) 

\item Jabs, A.: A conjecture concerning determinism, reduction, and measurement in quantum mechanics, arXiv:1204.0614 (2017) (Quantum Studies: Mathematics and Foundations {\bf3} (4), 278-292 (2016), DOI:10.1007/s40509-016-0077-7)

\item Einstein, A.: \"Uber einen die Erzeugung und Verwandlung des Lichtes betreffenden heuristischen Gesichtspunkt, Ann. Phys. (Leipzig) {\bf17}, 132-148 (1905). English translation in: Arons, A.B. and Peppard, M.B.: Einstein's proposal of the photon concept - a translation of the Annalen der Physik paper of 1905, Am. J. Phys. {\bf331} (5), 367-374 (1965)

\item Elitzur, A.C., Vaidman, L.: Quantum Mechanical Interaction-Free Measurements, Found. Phys. {\bf23} (7), 987-997 (1993)  

\item Summhammer, J., Rauch, H., Tuppinger, D.: Stochastic and deterministic absorption in neutron-interference experiments, Phys. Rev. A {\bf36} (9), 4447-4455 (1987)

\item Schr\"odinger, E.: Der erkenntnistheoretische Wert physikalischer Modellvorstellungen (Jahresbericht des Physikalischen Vereins zu Frankfurt am Main 1928/29, 1929) p. 44-51. English translation: Conceptual models in phsics and their philosophical value, in: Schr\"odinger, E.: Science, Theory and Man (Dover, New York, 1957) p. 148-165, especially p. 161, 162

\item For experimenal evidence of this see [1, Sec.~2.3]

\item Strekalov, D.V., Sergienko, A.V., Klyshko, D.N., Shih, Y.H.: Observation of Two-Photon ``Ghost'' Interference and Diffraction, Phys. Rev. Lett. {\bf74} (18), 3600-3603 (1995)

\item Yin, J., Cao, Y., Yong, H-L. et al.: Bounding the speed of `spooky action at a distance', arXiv:1303.0614 (Phys. Rev. Lett. {\bf110}, 260 407 (2013))

\item Guereiro, T., Sanguinetti, B., Zbinden, H., Gisin, N., Suarez, A.: Single-photon space-like antibunching, arXiv:1204.171

\item Hafner, M., Summhammer, J.: Experiment on interaction-free measurement in neutron interferometry, Phys. Letters A {\bf235}, 563-568 (1997) 

\item Kwiat, P., Weinfurter, H., Herzog, T. Zeilinger, A.: Interaction-Free Measurement, Phys. {\rm R}ev. Lett. {\bf74} (24), 4763-4766 (1995) 

\item Marchie van Voorthuysen, E.H. du: {\rm R}ealization of an interaction-free measurement of the presence of an object in a light beam, Am. J. Phys. {\bf64} (12), 1504-1507 (1996)   

\item Tsegaye, T., Goobar, E., Karlsson, A., Bj\"ork, G., Loh, M.Y., Lim, K.H.: Efficient interaction-free measurements in a high-finesse interferometer, Phys. Rev. A {\bf57} (6), 3987-3990 (1998)  

\item White, A.G., Mitchell, J.R., Nairz, O., Kwiat, P.G.: ``Interaction-Free'' Imaging, arXiv:quant-ph/9803060 (Phys. Rev. A {\bf58}, 605-608 (1998))

\item Kwiat, P.G., White, A.G., Mitchell, J.R., Nairz, O., Weihs, G., Weinfurter, A., Zeilinger, A.: High-efficiency quantum interrogation experiments via the quantum Zeno effect, arXiv:quant-ph/9909083 (Phys. Rev. Lett. {\bf83}, 4725-4728 (1999))  

\item Mirell, S., Mirell, D.: High Efficiency Interaction-free Measurement from Continuous Wave Multi-beam Interference, arXiv:quant-ph/9911076 

\item Namekata, N., Inoue, S.: High-efficiency interaction-free measurements using a stabilized Fabry-P\'{e}rot cavity, J. Phys. B: at. Mol. Opt. Phys {\bf39}, 3177-3183 (2006) 

\item Rauch, H., Werner, S.A.: Neutron Interferometry (Clarendon Press, Oxford, 2000); and literature quoted there 

\item Klein, A.G., Opat, G.I., Hamilton, W.A.: Longitudinal Coherence in Neutron Interferometry, Phys. Rev. Lett. {\bf50} (8), 563-565 (1983)  

\item Jabs, A., Ramos, R.: Comment on ``Longitudinal Coherence in Neutron Interferometry'', Phys. Rev. Lett. {\bf58} (1) 2274 (1987) 

\item Hong, C.K., Ou, Z.Y., Mandel, L.: Measurement of Subpicosecond Time Intervals between Two Photons by Interference, Phys. Rev. Lett. {\bf59} (18), 2044-2046 (1987)   

\item Mandel, L., Wolf, E.: Optical coherence and quantum optics (Cambridge University Press, Cambridge, 1995), p. 1087  

\item Dicke, R.H.: Interaction-free quantum measurements: A paradox?, 
Am. J. Phys {\bf49} (10), 925-930 (1981)   

\item Beth, R.A.: Mechanical Detection and Measurement of the Angular Momentum of Light, Phys. Rev. {\bf50}, 115-125 (1936) 

\item Nogues, G., Rauschenbeutel, A., Osnaghi, S., Brune, M., Raimond, J.M., Haroche, S.: Seeing a single photon without destroying it, Nature {\bf400} (issue 6741), 239-242 (1999) 

\item Wheeler, J.A.: The ``Past'' and the ``Delayed-Choice'' Double-Slit Experiment, in: Marlow, A.R. (ed.): Mathematical Foundations of Quantum Theory (Academic Press, New York, 1978), p. 9-48   

\item Wheeler, J.A.: Law Without Law, in: Wheeler, J.A., Zurek, W.H. (eds.): Quantum Theory and Measurement (Princeton University Press, Princeton, 1983), p. 182-213  

\item Kawai, T., Ebisawa, T., et 7 al.: Realization of a delayed choice experiment using a multilayer cold neutron pulser, Nuclear Instruments and Methods in Physics Resarch A {\bf410}, 259-263 (1998) 

\item Lawson-Daku, B.J., Asimov, R., et 4 al.: Delayed choices in atom Stern-Gerlach interferometry, Phys. Rev. A {\bf54} (6), 5042-5047 (1996)

\item  Jacques, V., Wu, E., Grosshans, F., Traussart, F., Aspect, A., Grangier, Ph., Roch, J.-F.: Experimental realization of Wheeler's delayed-choice GedankenExperiment, arXiv:quant-ph/0610241 (almost idenical with: Science {\bf315}, 966-968 (2007))  

\item  Kim, Y-H., Yu, R., Kulik, S.P., Shih, Y.H.: A Delayed Choice Quantum Eraser, arXiv:quant-ph/9903047 (Phys. Rev. Lett. {\bf84}, 1-5 (2000)) 

\item Baldzuhn, J., Mohler, E., Martienssen, W.: A wave-particle delayed-choice experiment with a single-photon state, Z. Phys. B - Condenced Matter {\bf77}, 347-352 (1989)   

\item Hellmuth, T., Walther, H., Zajonc, A., Schleich, W.: Delayed-choice experiments in quantum interference, Phys. Rev.A {\bf35} (6), 2532-2541 (1987)  

\item  Kaiser, F., Coudreau, T., Milman, P., Ostrowsky, D.B., Tanzilli, S.: Entanglement-enabled delayed choice experiment, arXiv:1206.4348 (Science {\bf338}, 637-640 (2012)) 

\item  Ma, X.-S., Kofler, J., Qarry, A., Tetik, N., Scheidl, T., Ursin, R., Ramelow, S., Herbst, T., Ratschbacher, L., Fedrizzi, A., Jennewein, T, Zeilinger, A.: Quantum eraser with causally disconnected choice, arXiv:1206.6578 (Proc. Natl. Acad. Sci. USA, {\bf110}, 1221-1226 (2013))

\item Peruzzo, A., Shadbolt, P., Brunner, N., Popescu, S., O'Brien, J.L.: A quantum delayed choice experiment, arXiv:1205.4926 (Science {\bf338} (6107), 634-637 (2012)) 

\item Kim, Y.-H., Yu, R., Kulik, S.P., Shih, Y., Scully, M.O.: Delayed ``Choice'' Quantum Eraser, Phys. Rev. Lett. {\bf84} (1), 1-5 (2000) 

\item Hasegawa, Y., Kikuta, S.: Apparent Destruction of Interference by Use of a Partial Absorber in X-Ray Interferometry, Japanese Journal of Applied Physics, {\bf30}, L316-L318 (1991) 

\item Awaya, K., Tomita, M.: Stochastic and deterministic absorption in single-photon interference experiments, Phys. Rev. {\bf56} (5), 4106-4110 (1997)

\item Muynck, W.M. de, Martens, H.: Neutron interferometry and the joint measurement of incompatible observables, Phys. Rev. A {\bf42} (9), 5079-5085 (1990)

\end{enumerate}
\bigskip
\hspace{5cm}
------------------------------
\end{document}